
\input amstex
\input amsppt.sty
\input crossref
\nologo
\pagewidth{160truemm}
\pageheight{247truemm}
\input epsf
\raggedbottom

\TagsOnRight

\def\Real{{\Bbb R}}
\def\sob#1#2#3{{\Vert#3\Vert}^{#1}_{#2}}
\def\sobsq#1#2#3{\bigl(\sob{#1}{#2}{#3})^2}
\def\norm#1{\sob{}{}{#1}}
\let\leq=\leqslant

\let\epsilon=\varepsilon
\def\cd{\Cal{D}}
\def\ssk{\noalign{\smallskip}}

\topmatter
\title
Generalised hyperbolicity in space-times with conical singularities
\endtitle
\author
J.~P.~Wilson
\endauthor
\address
Faculty of Mathematical Studies, University of Southampton, Highfield,
Southampton SO17 1BJ, UK.
\endaddress
\email
jpw@maths.soton.ac.uk
\endemail
\abstract
It is shown that the space-time with a conical singularity, which describes
a thin cosmic string, is hyperbolic in the sense that a unique $H^1$
solution exists to the initial value problem for the wave equation with a
certain class of initial data.
\endabstract
\endtopmatter

\document
\crossreferencing

\head\secnum. Introduction \endhead

A desirable property of any space-time used to model a physically plausible
scenario is that the evolution of the Einstein's equations is well posed
i.e.\ the initial value problem has a unique solution.  Space-times whose
metrics are at least $C^{2-}$, which guarantees the existence of unique
geodesics, fall within the context of the Cosmic Censorship Hypothesis of
Penrose~(1979). The hypothesis states that that the space-time will be
globally hyperbolic, i.e. strong causality is satisfied and $J^+(p)\cap
J^-(q)$ is compact $\forall p,\,q\in M$, and hence the evolution of
Einstein's equations is well defined.

There are however a number of space-times with weak singularities which
model physically plausible scenarios such as thin cosmic strings
(Vickers,~1987), impulsive gravitational waves (Penrose,~ 1972) and dust
caustic space-times (Clarke and O'Donnell,~1992). Typically such a
space-time has a locally bounded metric whose differentiability level is
lower than $C^{2-}$, but whose curvature is well defined as a distribution,
often with its support on a proper submanifold. Although cosmic censorship
may be violated for such space-times, it does not rule out the possibility
that the evolution of some fields is well posed.

A concept of hyperbolicity for such space-times was proposed by
Clarke~(1998). This was based on the extent to which singularities
disrupted the local evolution of the initial value problem for the scalar
wave equation.
$$ \aligned
   \square \phi & =f \\
   \phi_{|S} &= \phi_0 \\
   n^a {\phi_{,a}}_{|S} &= \phi_1
   \endaligned
$$
Clarke reformulated the initial value problem, on an open region $\Omega$
with a compact closure admitting a space-like hypersurface $S$, which
partitions $\Omega$ into two disjoint sets $\Omega^+$ and $\Omega^-$, in a
distributional form, obtained by multiplying a test field $\omega$ and
integrating by parts once to give
$$ \aligned
   &\int_{\Omega^+}\!\!\! \phi_{,a} \omega_{,b} g^{ab} \,(-g)^{1/2}\,d^4x
   = -\int_{\Omega^+}\!\!\!f \omega\,  (-g)^{1/2} \,d^4x
   - \int_{S} \phi_1 \omega \, dS \qquad \forall\omega\in\cd(\Omega) \\ 
   &\phi_{|S}=\phi_0
   \endaligned
$$
and then defined a point $p\in M$ as being {\it $\square$-regular} if it
admitted such a neighbourhood $\Omega$ for which the above equation had a
unique solution for each set of Cauchy data $(\phi_0,\phi_1)\in
H^1(S)\times H^0(S)$.  A space-time which was $\square$-regular everywhere
could then be said to be {\it $\square$-globally hyperbolic}.  It was shown
if a space-time satisfied the following curve integrability conditions at a
given point $p\in M$, then that point was $\square$-regular.
\roster
\item $g_{ab}$ and $g^{ab}$ are continuous
\item $g_{ab}$ is $C^1$ on $M-J^+(p)$
\item $g_{ab,c}$ exists as a distribution and is locally square integrable.
\item The distributional Riemann tensor components $R^{a}{}_{bcd}$ may be
interpreted as locally integrable functions
\item There exists a non empty open set $C\subset\Real^4$ and functions
$M,N:\Real^+\to\Real^+$, with $M(\epsilon),\,N(\epsilon)\to0$ as $\epsilon\to0$, such that if
$\gamma:[0,1]\to M$ is a curve with $\dot\gamma\in C$ then $\gamma$ is
future time-like and
$$ \aligned
   &\int^\epsilon_0 \,\bigl| \Gamma^a_{bc}(\gamma(s)) \bigr|^2\,ds <
   M(\epsilon) \\ 
   &\int^\epsilon_0 \,\bigl| R^{a}{}_{bcd}(\gamma(s))\bigr| \,ds < N(\epsilon)
   \endaligned
$$
\endroster
The proof involved the construction of a congruence of time-like geodesics,
whose tangent admitted an essentially bounded weak derivative, and a
suitable energy inequality from which uniqueness and existence could be
deduced. In particular it was shown that these results were applicable to
the dust caustic space-times.

Not all space-times with weak singularities satisfy these integrability
conditions. One such example is a space-time with a conical singularity
representing a thin string. Such a space-time may be constructed by cutting
out a sector of angle $2\pi(1-A)$ from Minkowski space-time and identifying
the resulting edges. In a suitable Cartesian coordinate system
$(t,x,y,z)\in\Real^4$ the metric $g_{ab}$ may be written as the following
line element
$$ ds^2=-dt^2+dx^2+dy^2+dz^2-{(1-A)^2\over x^2+y^2} (x\,dy-y\,dx)^2
   \tag\tagnum\xlabel{conical} $$
and the non-zero energy-momentum tensor components may be shown to be the
well defined distribution (Balasin and Nachbagauer, 1993; Clarke et al.,
1996)
$$ T^t{}_t (-g)^{1/2} = T^z{}_z (-g)^{1/2} = -2\pi(1-A)
   \delta(x)\delta(y)\tag\tagnum\xlabel{emt}$$ 
which justifies the interpretation of this space-time as representing a
thin cosmic string.  Curve integrability is violated at the axis because
$g_{ab,c}$ is not locally square integrable there, however there does exist
a natural congruence of time-like geodesics; namely the integral curves of
the vector $\partial/\partial t$ whose covariant derivative vanishes. In
this paper it will be shown that $\square$-regularity is also applicable to
such space-times.

\subhead
A remark on notation
\endsubhead

The lower case indices $a$, $b$ etc.\ denote coordinates $t$, $x$, $y$ and
$z$ where as the upper case indices $A$, $B$ etc.\ denote coordinates $x$
and $y$.

The Sobolev spaces $H^k(\Omega)$, where $\Omega$ is an open subset of
$\Real^n$, are defined as the spaces of functions $f\in L^2(\Omega)$ such
that all partial derivatives up to order $k$ are defined as distributions
and such that the Sobolev norm
$$ \sob{k}{\Omega}{f} = \biggl(\sum^k_{\alpha_1+\cdots+\alpha_n=0}
   \int_\Omega \,\, \Bigl|{ \partial^{\alpha_1+\cdots+\alpha_n} f \over 
   (\partial x^1)^{\alpha_1}\ldots (\partial x^n)^{\alpha^n}}\Bigr|^2
   \,d^nx \biggr)^{1/2} $$ 
is finite.

For an $m$-dimensional subspace $S\in\Omega$ we may also define the
embedding norm
$$ \sob{k}{S\subset\Omega}{f} = \biggl(\sum^k_{\alpha_1+\cdots+\alpha_n=0}
   \int_S \,\, \Bigl|{ \partial^{\alpha_1+\cdots+\alpha_k} f \over (\partial
   x^1)^{\alpha_1}\ldots (\partial x^n)^{\alpha^n}}\Bigr|^2 \,d^mx
   \biggr)^{1/2} $$ 
It should be noted that derivatives in all tangential directions of
$\Omega$ are summed over, not only those tangential to $S$. Clearly
$\sob{k}{S}{f}\leq\sob{k}{S\subset\Omega}{f}$.

\head\secnum. The initial value problem \endhead

In order to prove hyperbolicity, we must show that in each open region
$\Omega$ with a compact boundary $\partial \Omega$, that a unique solution
$\phi\in H^1(\Omega^+)$ exists to the initial value problem;
$$ \aligned
   \square \phi & =f \\
   \phi_{|S} &= \phi_0 \\
   n^a {\phi_{,a}}_{|S} &= \phi_1
   \endaligned
   \tag\tagnum\xlabel{ivp}
$$
where $f\in H^{0}(\Omega)$, $\phi_0\in H^{1}(S)$, $\phi_1\in H^1(S)$.

We shall assume that without loss of generality that $\Omega$ is generated
by a foliation of space-like hypersurfaces $(S_\tau)_{\tau_1<\tau<\tau_2}$
(with $\tau_1<0<\tau_2$) having a common boundary, with $S_0$ coinciding
with the initial hypersurface $S$ described by $t=0$. The region of
$\Omega$ to the future of $S$, $\Omega^+$ will be denoted as $U$.

\topinsert
\centerline{\epsfbox{foliation.ps}}
\bigskip
\bigskip
\centerline{Figure 1. The foliation of region $\Omega$}
\endinsert

There will now be two cases to consider, according to whether $U$
intersects the axis $\Lambda$ given by $x=y=0$. In the case where
$U\cap\Lambda=\emptyset$, the metric is equivalent to that of Minkowski
space and so one can write down a unique solution in terms of the Kirchoff
formula (see e.g.\ Choquet-Bruhat et al.,~1977). We shall therefore
consider the more problematic case $U\cap\Lambda\neq\emptyset$, where the
metric is locally bounded but admits directional dependent limits as one
approaches the axis.

By writing $\phi=\psi+q$ where $q\in H^1(U)$ is an arbitrary function
satisfying
$$ \aligned
   q_{|S} &=\phi_0 \\
   n^a {q_{,a}}_{|S} &=\phi_1
   \endaligned
   \tag\tagnum\xlabel{ivpq} $$
we can express~\eqref{ivp} as an initial value problem for $\psi$
satisfying zero initial conditions
$$ \aligned
   \square \psi & =f-\square q \\
   \psi_{|S} &= 0 \\
   n^a {\psi_{,a}}_{|S} &= 0
   \endaligned
   \tag\tagnum\xlabel{ivp2}
$$

One possible way of defining  a function $q\in L^2(U)$ which satisfies the
initial conditions~\eqref{ivpq} is
$$ \aligned
   &q(t,x,y,z)=\phi_0(x,y,z)+t\phi_1(x,y,z)-{t^3\over t^2+x^2+y^2} \bigl(
   x\phi_{1,x}(0,0,z)+y\phi_{1,y}(0,0,z) \bigr) \\
   &(\phi_0,\phi_1) \in H^1(S)\times H^0(S)
   \endaligned
$$
By imposing stronger initial conditions, we can also achieve $q\in
H^1(U)$. On differentiating
$$ \aligned
   q_{,t} &= \phi_1(x,y,z) - {t^2(t^2+3x^2+3y^2) \over (t^2+x^2+y^2)^2}
   \bigl(x\phi_{1,x}(0,0,z) +y\phi_{1,y}(0,0,z)\bigr)\\
   q_{,z} &= \phi_{0,z}(x,y,z) + t \phi_{1,z}(x,y,z) \\
   & \qquad - {t^3\over t^2+x^2+y^2} \bigl( x\phi_{1,xz}(0,0,z) +
   y\phi_{1,yz}(0,0,z) \bigr)\\\ssk 
   q_{,A} &= \phi_{0,A}(x,y,z) + t \phi_{1,A}(x,y,z) \\
   &\qquad - {t^3\over t^2+x^2+y^2} \phi_{1,A}(0,0,z) \\
   &\qquad + 2 {t^3 x_A\over (t^2+x^2+y^2)^2}
   \bigl(x\phi_{1,x}(0,0,z) + y\phi_{1,x}(0,0,z) \bigr)
\endaligned $$
we can see that sufficient conditions on the initial data are
$$ \phi_0,\,\phi_1\in H^1(S), \qquad
   {\phi_{1,A}}_{|\Lambda}\in H^1(S\cap\Lambda) $$
Since we shall be working exclusively with the Sobolev spaces $H^k$; it
would be very desirable for $\square q\in H^0(U)$ so that $\square\psi\in
H^0(U)$ in~\eqref{ivp2}. This will require a further strengthening of the
conditions on the initial data $(\phi_0,\phi_1)$. Now $\square q$ may be
expressed in the form
$$ \square q=g^{ab} q_{,ab} - {(1/A^2-1)\over(x^2+y^2)^{1/2}}
   \alpha^a q_{,a} $$
where
$$ \alpha^a = \bigl(0, x/(x^2+y^2)^{1/2}, y/(x^2+y^2)^{1/2},0\bigr) $$
The non-zero components $g^{tt}$, $g^{AB}$, $g^{zz}$ and $\alpha^A$ are
essentially bounded. Therefore it is sufficient to require that $q_{,tt}$,
$q_{,AB}$, $q_{,zz}$, $q_{,A}/(x^2+y^2)^{1/2}$ are all square integrable.

On differentiating we find that
$$ \alignat1
    q_{,tt} &= -2 {t(x^2+y^2)(3x^2+3y^2-t^2)\over(t^2+x^2+y^2)^{5/2}}\\
      &\qquad\times\biggl({x\over(x^2+y^2)^{1/2}} \phi_{1,x}(0,0,z)+  
      {y\over(x^2+y^2)^{1/2}} \phi_{1,y}(0,0,z)\biggr) \\\ssk
    q_{,zz} &= \phi_{0,zz}(x,y,z) + t \phi_{1,zz}(x,y,z) \\
      &\qquad - {t^3\over t^2+x^2+y^2} \bigl(x \phi_{1,xzz}(0,0,z) +
      y\phi_{1,yzz}(0,0,z)\bigr)\\\ssk 
    q_{,AB} &= \phi_{0,AB}(x,y,z) + t \phi_{1,AB}(x,y,z) \\
   &\qquad + 2 {t^3  \over (t^2+x^2+y^2)^2} \bigl(
     x_B \phi_{1,A}(0,0,z) + x_B \phi_{1,A}(0,0,z)\bigr) \\
   &\qquad +2 {t^3\over(t^2+x^2+y^2)^{3/2}} \biggl(\delta_{AB} -4
   {x_Ax_B\over t^2+x^2+y^2} \biggr) \\
   &\qquad\qquad\times \biggl({x\over (x^2+y^2)^{1/2}} \phi_{1,x}(0,0,z) 
   + {y\over (x^2+y^2)^{1/2}} \phi_{1,y}(0,0,z) \biggr)
\endalignat
$$
Also we apply the  mean value theorem (with $\xi,\,\eta\in[0,1]$) to
$q_{,A}$ to obtain  
$$ \split {q_{,A}\over(x^2+y^2)^{1/2}}
     &= {1\over (x^2+y^2)^{1/2}} \phi_{0,A}(0,0,z)
      + {t(x^2+y^2)^{1/2}\over t^2+x^2+y^2} \phi_{1,A}(0,0,z) \\
     &\qquad + {x\over (x^2+y^2)^{1/2}} \bigl(\phi_{0,Ax}(\xi x,0,z)
      + t \phi_{1,Ax}(\xi x,0,z)\bigr) \\
     &\qquad + {y\over (x^2+y^2)^{1/2}} \bigl(\phi_{0,Ay}(0,\eta y,z)  
      + t \phi_{1,Ay}(0,\eta y,z)\bigr)\\ 
     &\qquad + 2{x_A t^3\over (t^2+x^2+y^2)}\biggl( {x\over
       (x^2+y^2)^{1/2}}\phi_{1,x}(0,0,z) + {y\over (x^2+y^2)^{1/2}}
        \phi_{1,y}(0,0,z) \biggr) 
    \endsplit
$$
Sufficient conditions on the initial data for the integrability of each of
these components are
$$ \vcenter{\halign{$\displaystyle#$\hfil&
    \qquad\qquad$\displaystyle#$\hfil \cr
   q_{,tt}\in L^2(U) & {\phi_{1,A}}_{|\Lambda}\in L^2(S\cap\Lambda) \cr
   q_{,zz}\in L^2(U) & \phi_{0,zz},\,\phi_{1,zz}\in L^2(S),\ 
     {\phi_{1,zzA}}_{|\Lambda}\in L^2(S\cap\Lambda) \cr\ssk
   q_{,AB}\in L^2(U) & \phi_{0,AB},\,\phi_{1,AB}\in L^2(S),\
     {\phi_{1,A}}_{|\Lambda}\in L^2(S\cap\Lambda) \cr\ssk
   {q_{,A}\over(x^2+y^2)^{1/2}}\in L^2(U) &
     \vcenter{\halign{$#$\hfil\cr\phi_{0,A}=O((x^2+y^2)^n)\hbox{ for some
   $n>0$},\cr 
     \phi_{1,A},\,\phi_{0,AB},\,\phi_{1,AB}\in L^2(S)\cr}}\cr
}}$$
We therefore have sufficient conditions on the initial data
$(\phi_0,\phi_1)$, for $\square q\in L^2(U)$, of
\roster
\item $\phi_{0,A}=O((x^2+y^2)^n)$ for some $n>0$
\item ${\phi_{1,A}}_{|\Lambda}\in H^{2}(S\cap\Lambda)$
\item $\phi_0,\,\phi_1\in H^{2}(S)$
\endroster

\head\secnum. Existence and uniqueness \endhead

On obtaining a function $q\in H^1(U)$ which makes the right hand side
of~\eqref{ivp2} square integrable, we may proceed to prove existence of
a unique solution $\psi\in H^1(U)$ to~\eqref{ivp2}, and hence a unique
solution $\phi\in H^1(U)$ to~\eqref{ivp}.  The first step in establishing
such a result is to construct an energy inequality (Clarke,~1998; Hawking
and Ellis,~1973).

We define the following energy integral
$$ E(\tau) = \int_{S_\tau}\!\!\! S^{ab} t_a n_b \,(-g)^{1/2}\, d^3x $$
where $t^a$ is the tangent to a time-like congruence of geodesics (here we
can take $t_a=\delta^t_a$), $n_a$ is the normal to the surface $S_\tau$ and
$$ S^{ab} = (g^{ac}g^{bd} - \tfrac12 g^{ab}g^{cd})\psi_{,a} \psi_{,b} -
   \tfrac12 g^{ab} \psi^2 $$
It will turn out that estimating this energy integral easier than working
directly with the classical Sobolev norm
$\sob{1}{S_\tau\subset\Omega}{\psi}$. However $E(\tau)$ and
$\sobsq{1}{S_\tau\subset\Omega}{\psi}$ are equivalent in the sense that one
can find global positive constants $B_1$ and $B_2$ such that
$$ B_1 \sobsq{1}{S_\tau\subset\Omega}{\psi} \leq E(\tau) \leq B_2
   \sobsq{1}{S_\tau\subset\Omega}{\psi} \tag\tagnum\xlabel{sandwich} $$

The crucial step in obtaining an energy inequality is to apply Stokes'
theorem to the vector $S^{ab}t_b$ on the region
$$ U_\tau= \bigcup_{0<\tau'<\tau} S_{\tau'} $$
however caution must be exercised because of the possible lack of
differentiability of $g_{ab}$ and $\psi$ in a neighbourhood of the axis
$\Lambda$. We therefore apply Stokes' theorem to the region
$$ U_\tau^\epsilon = \{\, x\in U_\tau \,|\, x^2+y^2>\epsilon^2 \} $$
which has a boundary $\partial U_\tau^\epsilon$ consisting of
$$ \aligned
   S^\epsilon_\tau &= S_\tau \cap U^\epsilon \\
   S^\epsilon &= S \cap U^\epsilon \\
   W^\epsilon_\tau &= \{\, (t,x,y,z) \in U_\tau \,|\,
   x^2+y^2=\epsilon^2\,\}
   \endaligned
$$
and will consider the limit $\epsilon\to0$.

On applying Stokes' theorem to this region we obtain
$$ \int_{U_\tau^\epsilon}\!\!\! (S^{ab}t_a)_{;b}\, (-g)^{1/2}\, d^4x
   = \int_{S^\epsilon_\tau}\!\!\! S^{ab} t_a n_b  \,dS
   - \int_{S^\epsilon}\!\!\! S^{ab} t_a n_b \,dS
   - \int_{W^\epsilon_\tau}\!\!\! S^{ab} t_a n_b \,dW $$
Taking the limit as $\epsilon\to0$ and using the fact that $t_{a;b}=0$
$$ E(\tau)=E(0)+F(\tau)+\int_{U_\tau}\!\!\! S^{ab}{}_{;b} t_a
   \,(-g)^{1/2}\,d^4x $$ 
where
$$ F(\tau)= \lim_{\epsilon\to0} \int_{W^\epsilon_\tau}\!\!\! S^{ab} t_a n_b
   \,dW $$
The quantity $F(\tau)$ is the limiting flux integral, whose significance we
shall discus later; for the time being we shall assume that this quantity
exists. We have
$$ E(\tau) = E(0) + F(\tau) + \int_{U_\tau}\!\!\! \psi_{,t}
   (\psi-\square\psi) \,(-g)^{1/2}\,d^4x $$
which implies that
$$ E(\tau) \leq E(0) + |F(\tau)|
   + K \sobsq{0}{U_\tau}{\square\psi}
   + K \sobsq{1}{U_\tau}{\psi} $$
Using the fact that
$$ \sobsq{1}{U_\tau}{\psi}
   = \int^\tau_0 \sobsq{1}{S_{\tau'}\subset\Omega}{\psi} \,d\tau'  $$
and on applying~\eqref{sandwich} we have
$$ E(\tau) \leq E(0) + |F(\tau)|
   + K \sob{0}{U_\tau}{\square\psi}
   + (K/B_1) \int^\tau_0 \!\!\!E(\tau')\,d\tau' $$
This inequality may be solved by Gronwall's Lemma (See e.g.\ Abraham et
al.,~1988) to give
$$ E(\tau) \leq \Bigl( E(0) + |F(\tau)| + K
   \sobsq{0}{U_\tau}{\square\psi} \Bigr) e^{K/B_1\,\tau} $$
or equivalently
$$ \sobsq{1}{S_\tau}{\psi} \leq  {1\over B_1} \Bigl( B_2
   \sobsq{1}{S\subset\Omega}{\psi} + |F(\tau)| + K
   \sobsq{0}{U_\tau}{\square\psi} \Bigr) e^{K/B_1\,\tau} 
   \tag\tagnum $$
The first term in the right hand side is determined by the initial data
in~\eqref{ivp2} and will be zero in our case.

We first consider the issue of whether any such solution to \eqref{ivp2} is
unique; suppose that $\gamma$ is the difference of two such solutions of
\eqref{ivp2}, then it must be a solution of the initial value problem
$$ \aligned
   \square\gamma & =0 \\
   \gamma_{|S} &= 0 \\
   n^a {\gamma_{,a}}_{|S} &= 0
   \endaligned
$$
and so the corresponding energy inequality is
$$ \sobsq{1}{S_\tau\subset\Omega}{\gamma} \leq {1\over B_1} |F(\tau)|
   e^{K/B_1\,\tau} 
$$
A vanishing limiting flux term $F(\tau)$ will imply that $\gamma=0$ and
therefore any such solution to~\eqref{ivp2} would be unique. In the case of
a non-vanishing $F(\tau)$, it could be possible for uniqueness to be
violated. However $F(\tau)$ may be expressed as
$$ F(\tau) = - \lim_{\epsilon\to0} \int_{W^\epsilon_\tau}\!\!\! \psi_{,t}
   (n^x \psi_{,x} + n^y \psi_{,y} ) \,dW \tag\tagnum\xlabel{flux} $$
Since the metric is locally bounded and the surface element can be
expressed as $A\epsilon\,dt\,d\varphi\,dz$ in suitable cylindrical
coordinates, a sufficient condition for $F(\tau)$ to vanish is that $\psi$
has a locally bounded derivative.

We finally consider the question of whether a solution exists
to~\eqref{ivp2} using a method following those of Egorov and Shubin~(1992)
and Clarke~(1998). We shall show that a $C^1$ solution exists
to~\eqref{ivp2}, whose differentiability is sufficient for the flux
term~\eqref{flux} to vanish and therefore guarantees that the solution is
unique. We first define the solution function space $V_0$ and its dual
$V_1$
$$ \aligned
   V_0 &= \{\, \psi\in C^1(U) \,|\, \psi_{|S}=n^a{\psi_{,a}}_{|S}=0 \,\} \\
   V_1 &= \{\, \omega\in C^1(U) \,|\, \omega_{|S_{\tau_2}}=
n^a{\omega_{,a}}_{|S_{\tau_2}}=0 \,\} \endaligned $$
We may apply the energy inequality to $\psi\in V_0$.  It should be noted,
as in the previous section, the imposed differentiability level will force
the flux term $F(\tau)$ to vanish and so we are left with
$$ \sobsq{1}{S_\tau\subset\Omega}{\psi} \leq B_2 K e^{K/B_1\,\tau_2}
   \sobsq{0}{U_\tau}{\square\psi}  $$
In particular we apply this to the region $U$ and integrate to obtain
$$ \norm{\psi} \leq c_1 \norm{\square\psi} \qquad \forall\psi\in V_0 $$
where we use $\norm\psi$ to denote the $L^2$ norm of $\psi$ over $U$.

We may obtain a similar inequality for $\omega\in V_1$, by regarding
$S_{\tau_2}$ as an initial surface, with zero initial data, evolving back
in time and constructing an analogous energy inequality; thus we obtain
$$ \norm{\omega} \leq c_2 \norm{\square\omega} \qquad \forall\omega\in V_1
   \tag\tagnum\xlabel{boxomega} $$

We now apply Stokes' theorem to $\psi\omega^{;a}$ and $\omega\psi^{;a}$ for
$\psi\in V_0$ and $\omega\in V_1$ noting that the boundary contributions on
$S$ and $S_{\tau_2}$ vanish.
$$ \aligned
   \int_{U^\epsilon}\!\!\! \psi^{;a}\omega_{;a}\,(-g)^{1/2}\, d^4x
    + \int_{U^\epsilon}\!\!\! \psi \square\omega\, (-g)^{1/2}\,d^4x
   &= - \int_{W^\epsilon_{\tau_2}}\!\!\! \psi \omega_{,a} n^a \,dW \\
   \int_{U^\epsilon}\!\!\! \psi^{;a}\omega_{;a}\,(-g)^{1/2}\, d^4x
    + \int_{U^\epsilon}\!\!\! \square\psi\,\omega\,(-g)^{1/2}\,d^4x
   &= - \int_{W^\epsilon_{\tau_2}}\!\!\!\omega \psi_{,a} n^a\, dW \\
\endaligned $$
On taking the limit $\epsilon\to0$, the flux integrals on the right hand
side vanish because $\psi$ and $\omega$ are $C^1$. Subtracting the
resulting equations gives
$$ \int_{U} \square\psi\,\omega (-g)^{1/2}\,d^4x = \int_{U} \psi
   \square\omega (-g)^{1/2}\,d^4x$$
which, because $(-g)^{1/2}$ is  constant may be written as
$$ \int_{U} \square\psi\,\omega d^4x = \int_{U} \psi
   \square\omega d^4x \tag\tagnum\xlabel{switchbox}$$
As a consequence, we may define a linear functional $k:\square V_1\to\Real$ by
$$ k(\square\omega)=\int_U(f-\square q)\omega d^4x $$
and by~\eqref{boxomega}
$$ \split
   \left| k(\square\omega) \right|
   &\leq \norm{f-\square q}\norm\omega \\
   &\leq c_2 \norm{f-\square q}\norm{\square\omega}
   \endsplit
$$
implying that this linear functional is also bounded.

In order for $k$ to correspond to an element of $V_0$ (a solution of
\eqref{ivp2}) we must show that $\square V_1$ is a dense subspace of
$L^2(U)$. It is sufficient to show that any function in the space $(\square
V_1)^\perp$ is necessarily zero.  Suppose that $\lambda\in (\square
V_1)^\perp$ then, because $V_0$ is dense, there exists a sequence
$(\psi_n)$ in $V_0$ converging to $\lambda$. By~\eqref{switchbox}, we have
for all $\omega\in V_1$
$$ \split
   \lim_{n\to\infty}\int_U \square\psi_n \omega \,d^4x
    &  = \lim_{n\to\infty} \int_U \square\omega \psi_n \,d^4x\\ 
    & = \int_U \square\omega \lambda \,d^4x \\
    & =0
    \endsplit
$$
Since $V_1$ is dense, this implies that $\square\psi_n\to0$ a.e.\ That is we
have $\lambda=0$

\head\secnum. Conclusion \endhead

We have shown that by using the functional analytic methods of
Clarke~(1998), that for initial data satisfying the following conditions; 
\roster
\item $\phi_{0,A}=O((x^2+y^2)^n)$ for some $n>0$
\item ${\phi_{1,A}}_{|\Lambda}\in H^{2}(S\cap\Lambda)$
\item $\phi_0,\,\phi_1\in H^{2}(S)$
\endroster
\noindent
a unique solution $\phi\in H^1(U)$ to~\eqref{ivp} exists in every open
region $U$ of the conical space-time~\eqref{conical}, and therefore such a
space-time can be regarded as hyperbolic in this sense.

The integrability conditions on the initial data resulted from the choice
of the arbitrary function $q\in H^1(U)$ which satisfied the initial
data. It may be possible to weaken them by a more careful choice of such a
function.  The actual proof of existence and uniqueness depended on two
properties of the metric; namely that both its covariant and contravariant
components are locally essentially bounded, and that there existed a
natural congruence of time-like geodesics whose tangent has an essentially
bounded covariant derivative.  It is therefore possible to modify the proof
to show that a similar evolution of the wave equation is possible in other
conical space-times whose metric differs from~\eqref{conical} by a $C^2$
perturbation.  In such a space-time the conical singularity will have a
constant deficit angle and the axis may be regarded as a totally geodesic
two dimensional time-like submanifold, and therefore represents a thin
string on a curved background.

More recently the same problem was approached by Vickers and Wilson~(1999)
using Colombeau's generalised functions (See e.g.\
Colombeau,~1984). Colombeau's theory enables a distributional 
interpretation to be given to products of distributions that would
otherwise not be defined in classical distribution theory, and therefore
has a natural application to non-linear theories such as General
Relativity. One important recent application was to rigorously establish
the form of the energy momentum tensor~\eqref{emt}.  In that paper, the
initial value problem~\eqref{ivp} was formulated as an initial value
problem in the Colombeau Algebra and existence and uniqueness was proved
within the algebra.  It should be noted that, by working with regularised
functions in a Colombeau Algebra, it was possible to establish existence
and uniqueness given {\it any} initial data $(\phi_0,\phi_1)\in
H^1(S)\times H^0(S)$, where as in this paper we needed stronger
differentiability conditions on the initial data.

\head Acknowledgements \endhead

The author wishes to thank Professor Chris Clarke and Dr James Vickers for
many helpful discussions.

\enddocument

\bye